\documentstyle[floats,twocolumn,aps,epsf]{revtex}

\begin{document}

\draft
\twocolumn[\hsize\textwidth\columnwidth\hsize\csname@twocolumnfalse\endcsname
\title{
Reduction of the Superfluid Density in the Vortex-Liquid Phase of ${\bf 
Bi_2Sr_2CaCu_2O_y}$
}
\author{
Tetsuo Hanaguri$^1$~\cite{hana}, Takashi Tsuboi$^1$, Yoshishige Tsuchiya$^1$, 
Ken-ichi Sasaki$^1$, and Atsutaka Maeda$^{1,2}$
}
\address{
$^1$Department of Basic Science, The University of Tokyo,\\
3-8-1, Komaba, Meguro-ku, Tokyo 153-8902, Japan
}
\address{
$^2$CREST, Japan Science and Technology Corporation (JST),\\
4-1-8, Honcho, Kawaguchi 332-0012, Japan
}
\date{\today}
\maketitle
\begin{abstract}
In-plane complex surface impedance of a ${\rm Bi_2Sr_2CaCu_2O_y}$ single 
crystal was measured in the mixed state at 40.8~GHz.
The surface reactance, which is proportional to the real part of the effective 
penetration depth, increased rapidly just above the first-order vortex-lattice 
melting transition field and the second magnetization peak field.
This increase is ascribed to the decrease in the superfluid density rather 
than the loss of pinning.
This result indicates that the vortex melting transition changes the 
electronic structure as well as the vortex structure.
\end{abstract}

]

\narrowtext
The electronic structure of high-$T_c$ superconductors (HTSC's) in the mixed 
state attracts much attention.
In conventional superconductors (CSC's) with $s$-wave gap, quasiparticle (QP)
states in the mixed state are localized in the vortex cores where the 
superconducting gap is suppressed.
The QP excitation spectrum inside the vortex core consists of quantized 
energy levels separated by $\Delta E \sim \Delta_0/k_F\xi$, where $\Delta_0$ 
is the bulk gap, $k_F$ is the Fermi momentum and $\xi$ is the coherence 
length~\cite{Caroli}.
Since $\Delta E$ in CSC's is much smaller than the scattering energies, the 
vortex core can be regarded as a normal metal.
In HTSC's, above picture can not be applied because of the following reasons.
First, the symmetry of the superconducting gap in HTSC's is not an $s$-wave 
but is most likely a $d$-wave .
Since the amplitude of the $d$-wave gap is zero at the node, QP's are not 
localized in the vortex core but extend along the node 
direction~\cite{Ichioka,Volovik}.
Calculations based on the Bogoliubov-de Gennes equations suggest that there 
are no truly localized states in the vortex core of {\it pure} $d$-wave 
superconductors~\cite{Wang,Himeda,Franz}.
Secondly, even if there are localized QP states inside the vortex core owing 
to the mixing of the different gap symmetry~\cite{Himeda,Franz,Laughlin}, 
$\Delta E$ may exceed any other energy scales, since $k_F\xi$ of HTSC's is 
very small.
Therefore the vortex core in HTSC's should be different from that of CSC's in 
either cases.
Finally, in HTSC's, the greater part of the mixed state is a vortex liquid 
phase which is practically lost in CSC's.
It may be possible that the difference in the vortex structure brings the 
difference in the electronic state.

Experimentally, scanning tunneling spectroscopies~\cite{Maggio,Renner} and 
thermal conductivity measurements~\cite{Krishana} have been used to study the 
QP state of HTSC's in the mixed states.
High frequency surface impedance $Z_s=R_s+iX_s$ also provides useful 
information on the electronic state.
Using high enough frequencies, one can be free from the vortex pinning and the 
information not only on QP's but also on superfluids can be deduced from $R_s$ 
and $X_s$, respectively.
So far, a vortex flow resistivity and a QP lifetime inside the vortex core 
were discussed from the $R_s$ measurements in 
${\rm YBa_2Cu_3O_y}$~\cite{Owliaei,Matsuda}.
Recently, Mallozzi {\it et al.} measured the complex resistivity in the mixed 
state of ${\rm Bi_2Sr_2CaCu_2O_y}$ and argued a possible $d$-wave 
effect~\cite{Mallozzi}.
However, measurements of $Z_s$ near the vortex-melting transition are still 
lacking.
In this paper, we report both $R_s$ and $X_s$ measurements on a 
${\rm Bi_2Sr_2CaCu_2O_y}$ single crystal and show that the high-frequency 
response can not be explained in terms of simple vortex flow.
Moreover, we succeeded in measuring $Z_s$ across the first-order vortex 
melting transition line and found that the reactive part, $X_s$, increases 
rapidly above the transition while there was little change in $R_s$.
This result indicates that the superfluid density or the {\it amplitude} of 
the order parameter decreases in the vortex liquid phase.

A ${\rm Bi_2Sr_2CaCu_2O_y}$ single crystal was grown by the floating zone 
method.
The as grown crystal was annealed in air at 800~${\rm ^{\circ}C}$ for 3 days 
and was quenched to room temperature to achieve an optimum oxygen content.
A superconducting transition temperature $T_c$ defined at zero resistivity was 
91~K.
Prior to the surface impedance measurements, we measured the magnetic-field 
dependence of the local magnetization of the same crystal using the micro Hall 
probe magnetometry and determined the first-order vortex melting 
transition field from the position of the magnetization jump.
Surface impedance was measured by the cavity perturbation technique with a 
cylindrical Cu cavity operated at 40.8~GHz in the TE$_{011}$ mode.
The sample was located at the center of the cavity which is the anti-node of 
the microwave magnetic field $H_{rf}$ being parallel to the $c$-axis of the 
sample.
The dimensions of the sample were 0.5$\times$0.5$\times$0.02~mm$^3$, which 
is appreciably larger than the normal-state skin depth ($\sim$1~$\mu$m at 
40.8~GHz).
The surface resistance $R_s$ and the surface reactance $X_s$ can be obtained 
from the changes in the quality factor of the cavity and the resonance 
frequency, respectively.
We determined absolute values of $R_s$ and $X_s$ from comparison with the 
dc resistivity and assuming that $R_s=X_s$ (Hagen-Rubens limit) in the normal 
state.
Using this procedure, we obtained the reasonable zero-temperature penetration 
depth of $\sim$2$\times$10$^2$~nm.
Surface impedance measurements were performed with swept temperature $T$ under 
field cooled conditions to avoid any extrinsic effects associated with pinning 
{\it e.g.} giant magnetostriction~\cite{Ikuta}.
In all the measurements, the static magnetic fields $H_{dc}$ were applied 
along the $c$-axis.

First, we briefly introduce the general behavior of the surface impedance of 
type-II superconductors~\cite{Coffey}.
The surface impedance $Z_s$ is related to the complex effective penetration 
depth $\tilde{\lambda}$ as $Z_s=i\mu_0\omega\tilde{\lambda}$, where $\mu_0$ is 
the vacuum permeability and $\omega$ is the angular frequency.
The complex resistivity $\tilde{\rho}$ is also expressed by $\tilde{\lambda}$ 
as $\tilde{\rho}=i\mu_0\omega\tilde{\lambda}^2$.
In the Meissner state, the response is reactive and $\tilde{\lambda}$ is 
purely real.
Therefore, $R_s\sim0$ and $X_s=\mu_0\omega\lambda_L$ where $\lambda_L$ is the 
London depth.
In the normal state, the response is dissipative and $\tilde{\lambda}^2$ is 
purely imaginary.
Therefore, $R_s=X_s=\mu_0\omega\delta/2$ where 
$\delta=(2\rho_n/\mu_0\omega)^{1/2}$ is the skin depth and $\rho_n$ is the 
normal resistivity.
In the mixed state, vortex dynamics affects $\tilde{\lambda}$.
If the frequency is low, vortices are effectively pinned and a response is 
similar to that of the Meissner state except that $\tilde{\lambda}^2\sim 
\lambda_L^2+B\Phi_0/\mu_0\kappa_p$, where $B$ is the magnetic induction, 
$\phi_0$ is the flux quantum and $\kappa_p$ is the Labusch parameter which 
denotes the pinning strength.
On the other hand, if the frequency is high enough, the viscous loss becomes 
dominant and a response is similar to that of the normal state except that 
$\delta$ is replaced by the vortex flow skin depth $\delta_f\sim 
(2B\Phi_0/\mu_0\omega\eta)^{1/2}$, where $\eta$ is the viscosity of the vortex 
motion being related to the QP excitation inside the vortex core.
Within the vortex flow theory of Bardeen and Stephen~\cite{Bardeen}, $\eta$ is 
independent of $B$ and $R_s\sim X_s\propto B^{1/2}$ at high enough fields 
where $\delta_f \gg \lambda_L$ and $R_s\propto B$, $X_s\sim 
\mu_0\omega\lambda_L$ at low fields.
A crossover from the pinned regime to the dissipative regime occurs at the 
pinning frequency $\omega_p=\kappa_p/\eta$.
\begin{figure}
\leavevmode\epsfxsize=7cm
\epsfbox{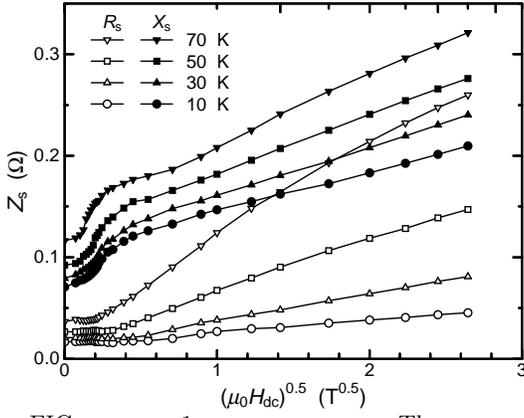}
\caption{
The complex surface impedance in ${\rm Bi_2Sr_2CaCu_2O_y}$ v.s. square-root of the applied field $H_{dc}$.
The surface reactance $X_s$ is larger than the surface resistance $R_s$, 
indicating that the response is reactive rather than dissipative.
Note that a prominent increase in $X_s$ at low fields.
}
\end{figure}

The field dependence of $Z_s$ of the ${\rm Bi_2Sr_2CaCu_2O_y}$ single crystal 
is shown in Fig.~1.
At high fields above 0.2~T, $R_s$ is almost proportional to $B^{1/2}$.
However, $X_s$ is larger than $R_s$ up to the highest field we used.
This result means that the contribution from the reactive parts, superfluid 
and/or pinned vortices, can not be neglected.
Therefore, the observed $B^{1/2}$ dependence of $Z_s$ can not be attributed to 
the high-field region of the Bardeen-Stephen type vortex flow.
\begin{figure}
\leavevmode\epsfxsize=7cm
\epsfbox{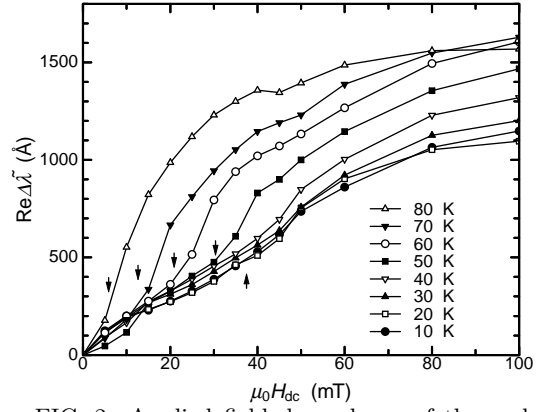}
\caption{
Applied field dependence of the real part of the effective penetration depth.
Downward arrows show the vortex melting transition field determined by the magnetization measurement. The upward arrow shows the second magnetization peak 
field at 30~K.
}
\end{figure}

At low fields, a prominent anomaly can be seen.
Above a certain field, $X_s$ increases rapidly and saturates at a higher field 
while there is little change in $R_s$ in the same field region.
To examine this anomaly in more detail, we plot the field induced changes in 
the real part of the effective penetration depth 
Re$\Delta\tilde{\lambda}(H_{dc})\equiv (X_s(H_{dc})-X_s(0))/\mu_0\omega$ in 
Fig.~2.
At low fields and low temperatures, Re$\Delta\tilde{\lambda}$ is independent 
of $T$ and linear in $H_{dc}$.
At high temperatures, Re$\Delta\tilde{\lambda}$ begins to deviate from the 
linear $H_{dc}$ dependence and rapidly increases above a certain field 
$H_{kink}$.
In Fig.~3, $H_{kink}$ is plotted on the magnetic phase diagram together with 
the vortex melting transition field $H_m$ and the second magnetization peak 
field $B_{sp}$.
Here, $H_{kink}$ is defined at the onset of the rapid increase in 
Re$\Delta\tilde{\lambda}(T)$ as shown in the inset of Fig.~3.
The agreement between $H_m$ and $H_{kink}$ is excellent, indicating that the 
vortex melting affects Re$\Delta\tilde{\lambda}$.
In the present crystal, the ``critical point'' is located at 45~K and 33~mT.
Above this field, the sharp feature in Re$\Delta\tilde{\lambda}(T)$ disappears 
as shown in the inset of Fig.~3.
However, as shown in Fig.~2, the small deviation from the linear $H_{dc}$ 
dependence is still observed in Re$\Delta\tilde{\lambda}(H_{dc})$ around 
40~mT, somewhat higher than $B_{sp}$.
Considering the difference between the applied field and the magnetic 
induction inside the sample and the different processes of the field 
applications in the local magnetization and the microwave measurements, it is 
reasonable to regard that the increase in Re$\Delta\tilde{\lambda}(H_{dc})$ 
below 45~K and the second magnetization peak have the same origin, most likely 
the field induced vortex decoupling~\cite{Bernhard}.
To sum up, Re$\Delta\tilde{\lambda}(H_{dc})$ increases when the 3-dimensional 
vortex lattice model is no longer valid.
Note that this anomaly only appears in the reactive response 
(Re$\Delta\tilde{\lambda}$ or $X_s$) and almost no anomaly appears in the 
dissipative response ($R_s$).

Now we discuss the origin of the change in Re$\Delta\tilde{\lambda}$.
Since ${\rm Bi_2Sr_2CaCu_2O_y}$ does not contain magnetic ions, we can 
recognize that Re$\Delta\tilde{\lambda}$ consists of two components: the 
London penetration depth $\lambda_L$ and the vortex motion.
First we interpret our results in terms of the change in the vortex motion, 
which has been widely assumed so far~\cite{Golosovsky}, and will show that 
this scenario is inappropriate at least near the vortex melting transition of 
${\rm Bi_2Sr_2CaCu_2O_y}$.
From the vortex-motion point of view, the increase in 
Re$\Delta\tilde{\lambda}$ should be attributed to the loss of pinning at the 
vortex melting transition.
Since contributions from $\lambda_L$ and the vortex motion can be separated if 
the response is written by the complex resistivity~\cite{Coffey}.
When the effect of vortex creep, which is important only near $T_c$, is 
neglected, the real ($\rho_1$) and the imaginary ($\rho_2$) parts of the 
complex resistivity are given by:

\begin{mathletters}
\label{rho}
\begin{equation}
\label{rho_1}
\frac{\rho_1}{\mu_0\omega\lambda_L^2}=\frac{s}{1+s^2}+\frac{r^2+sr}{(1+s^2)(1+r^2)}b,
\end{equation}
\begin{equation}
\label{rho2}
\frac{\rho_2}{\mu_0\omega\lambda_L^2}=\frac{1}{1+s^2}+\frac{r-sr^2}{(1+s^2)(1+r^2)}b.
\end{equation}
\end{mathletters}
\begin{figure}
\leavevmode\epsfxsize=7cm
\epsfbox{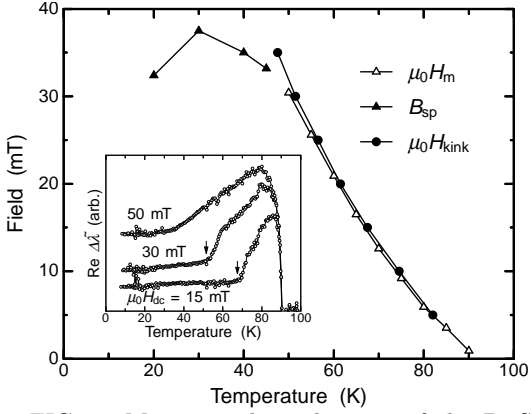}
\caption{
Magnetic phase diagram of the ${\rm Bi_2Sr_2CaCu_2O_y}$ single crystal used in 
this study.
The kink field $H_{kink}$ in Re$\Delta\tilde{\lambda}$, which is determined 
from the temperature dependence data as shown in the inset, exactly coincide 
with the vortex melting transition field $H_m$.
As for the second magnetization peak field, we plot the local field $B_{sp}$ 
instead of the applied field $H_{sp}$ since the self-field becomes large at 
low temperatures.
Above 50~K, the difference between the applied field and the local field is 
less than 1~mT.
}
\end{figure}

\noindent
Here, $s=(2\lambda_L^2/\delta_{nf}^2)$, $\delta_{nf}$ is the normal-fluid skin 
depth, $r=\omega/\omega_p$ and $b=B\phi_0/\mu_0\omega\eta\lambda_L^2$. 
The parameters $s$ and $r$ denote the normal-fluid fraction relative to the 
superfluid density and the weakness of the pinning, respectively.
Note that all the parameters related to the vortex motion ($r$ and $b$) are 
included only in the second terms of the right hand sides of Eq.~\ref{rho}.
Therefore, if only the vortex motion is considered, magnetic field dependent 
parts of the complex resistivity $\Delta\rho_1$, $\Delta\rho_2$ are nothing 
but these terms.
Here, we introduce the ratio of these two terms:
$\Delta\rho_2/\Delta\rho_1=(r-sr^2)/(r^2+sr)$.
Let us consider the behavior of this ratio at the vortex melting transition.
If the vortices melt, the effective pinning strength must be weakened.
Accordingly, $r\propto 1/\kappa_p$ should increase above the transition.
Since $\Delta\rho_2/\Delta\rho_1$ monotonically decreases with increasing $r$, 
$\Delta\rho_2$ should decrease relative to $\Delta\rho_1$ at the transition.
This behavior is expected to be irrespective of the nature of pinning, surface 
or bulk, because any depinning processes cause energy losses which mainly 
affect $\rho_1$.
To compare the above arguments with the experimental results, we calculated 
$\tilde{\rho}$ from $Z_s$.
As shown in Fig.~4, the experimentally obtained $\Delta\rho_2$ {\it increases} 
relative to $\Delta\rho_1$ at the vortex melting transition and the second 
magnetization peak field.
Therefore, the increase in Re$\Delta\tilde{\lambda}$ can not be attributed to 
the loss of pinning.
Instead, it should be originated from the increase in $\lambda_L$ itself.
Since $1/\lambda_L^2$ is proportional to the superfluid density, this result 
strongly suggest that the superfluid density decreases in the vortex liquid 
phase of ${\rm Bi_2Sr_2CaCu_2O_y}$ due to the additional pair breaking.
If all the QP's are localized in the vortex core as in $s$-wave 
superconductors, the electronic state might be insensitive to the change in 
the vortex structure.
Therefore, we speculate that the reduction of the superfluid density at the 
vortex melting transition is related to the $d$-wave superconductivity.
\begin{figure}
\leavevmode\epsfxsize=7cm
\epsfbox{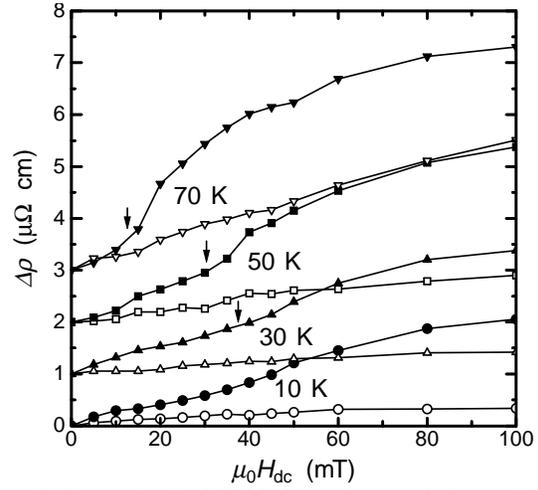}
\caption{
Applied field dependence of the real ($\Delta\rho_1$: open symbols) and 
imaginary ($\Delta\rho_2$: closed symbols) parts of the field dependent part 
of the complex resistivity calculated from $Z_s$.
Origins are shifted by 1~$\mu\Omega$~cm for each temperature.
Arrows show the vortex melting transition fields (70~K and 50~K) and the 
second magnetization peak field (30~K).
Note that difference between $\Delta\rho_2$ and $\Delta\rho_1$ increases above 
the transition. 
}
\end{figure}

Next we discuss the behavior of Re$\Delta\tilde{\lambda}$ in the vortex solid 
phase.
Since $R_s\sim 0$ in this regime, we can recognize that vortices are 
effectively pinned so that $r\ll 1$, and $\tilde{\lambda}\sim 
(\lambda_L^2+B\Phi_0/\mu_0\kappa_p)^{1/2}\sim 
\lambda_L+B\Phi_0/2\lambda_L\mu_0\kappa_p$.
If $\lambda_L$ is independent of $H_{dc}$, $\kappa_p$ should also be 
independent of $H_{dc}$ and $T$, since Re$\Delta\tilde{\lambda}\propto B$ and 
independent of $T$ as shown in Fig.~2.
Field independent $\kappa_p$ is only expected when the vortices are 
individually pinned by strong pinning centers.
This is unreasonable because we observed the sharp first-order magnetization 
jump in our sample.
Therefore, $\lambda_L$ as well as $\kappa_p$ should be $H_{dc}$ dependent even 
in the vortex solid phase.
Let us assume that the field dependence of $\lambda_L$ is dominant, as has 
been argued by Mallozzi {\it et al.} in the vortex liquid 
phase~\cite{Mallozzi}.
The field dependent $\lambda_L$ in the mixed state of $d$-wave superconductors 
is discussed in terms of the magnetic-field effect on the extended 
QP state by introducing the Doppler energy shift due to the 
circulating current around the vortices~\cite{Mallozzi,Kuebert,Vekhter}.
According to these theories, normal-fluid density, which is proportional to 
the change in $\lambda_L$, varies as $H_{dc}^{1/2}$ when $E_H>\Gamma$ and 
varies as linear in $H_{dc}$ when $E_H<\Gamma$, where $E_H$ is the averaged 
QP energy shift given by $E_H\sim \Delta_0(H_{dc}/H_{c2})^{1/2}$ 
and $\Gamma$ is the QP scattering rate by impurities or the thermal 
agitation.
In terms of this scenario, since the observed Re$\Delta\tilde{\lambda}$ in the 
vortex solid phase is linear in $H_{dc}$ and {\it independent} of $T$, 
$E_H<\Gamma$ and $\Gamma$ should be dominated by the impurity scattering.
However, the zero-field penetration depth of the same sample is found to be 
linear in $T$ at low temperatures.
This suggests that the dominant pair breaking mechanism should not be the 
impurity scattering but the thermal smearing.
To resolve the discrepancy between Re$\Delta\tilde{\lambda}$ in the vortex 
solid phase and the zero-field $\lambda_L$, knowledge of the field 
dependence of $\kappa_p$ and/or more detailed theoretical treatment must be 
needed.

Finally, we should briefly consider Re$\Delta\tilde{\lambda}$ above $H_m$.
As shown in Fig.~2, change in Re$\Delta\tilde{\lambda}$ just above the 
transition is rather gradual one.
Considering that the melting transition is of the first order, there should be 
a small discontinuity in Re$\Delta\tilde{\lambda}$ at $H_m$, which, we 
believe, is not observed because of sparse data points.
The succeeding gradual change at higher fields may come from the increasing 
pair breaking and/or the vortex motion.
If all the changes in Re$\Delta\tilde{\lambda}$ in the vortex liquid 
phase arise from the pair breaking~\cite{Mallozzi}, the quantity 
$\lambda_L(T=0)^2/({\rm Re}\tilde{\lambda})^2$ at low temperatures represents 
the superfluid fraction $f_s$.
We calculated this quantity at 10~K and found that 
$\lambda_L(T=0)^2/({\rm Re}\tilde{\lambda})^2\sim0.3$ at $H_{dc}$=1~T.
Even if $d$-wave effect is taken into account, this value is too small since 
$f_s$ is given by $f_s(H_{dc})\sim 1-(H_{dc}/H_{c2})^{1/2}$\cite{Vekhter} and 
$H_{c2}$ is considered to be around 100~T.
Therefore, both the increase in $\lambda_L$ and the contribution from the 
vortex motion should be considered in the vortex liquid phase.
We stress here again that the behavior of $\Delta\rho_2/\Delta\rho_1$ at the 
transition can not be explained in terms of the vortex motion alone, even 
though the two effects contribute to Re$\Delta\tilde{\lambda}$.
To separate both contributions, detailed frequency dependence measurements 
are indispensable.
This experiment is now underway.

In conclusion, we have measured the complex surface impedance $Z_s=R_s+iX_s$ 
of a ${\rm Bi_2Sr_2CaCu_2O_y}$ single crystal in the mixed state.
We succeeded in detecting the change in $Z_s$ at the first-order vortex 
melting transition and the second magnetization peak field for the first time.
Above the transition, $X_s$ which is proportional to the real part of the 
effective penetration depth increases while there was little change in $R_s$.
From the analysis of the complex resistivity deduced from $Z_s$, we showed 
that the increase in $X_s$ can not be ascribed to the loss of pinning but 
arises from the reduction of the superfluid density.
Namely, the additional pair breaking mechanism may exist in the vortex liquid 
phase.
Our results indicate that not only the phase but also the amplitude of the 
order parameter take different values in different vortex phases.
We speculate that this effect is related to the $d$-wave superconductivity.

The authors thank to Y. Matsuda and H. Kitano for helpful discussions.
This work was partly supported by the Grant-in-Aid for Scientific Research 
from the Ministry of Education, Science, Sports and Culture of Japan.

\end{document}